# Exploring the impact of under-reported cases on the COVID-19 spatiotemporal distributions using healthcare workers infection data

**Peixiao Wang [1] , Tao Hu [2,3,*], Hongqiang Liu [4] and Xinyan Zhu [1,5,6]**

1 State Key Laboratory of Information Engineering in Surveying, Mapping and Remote Sensing, Wuhan University, Wuhan 430079, China; peixiaowang@whu.edu.cn; xinyanzhu@whu.edu.cn
2 Department of Geography, Oklahoma State University, OK 74078, USA; taohu@g.harvard.edu
3 Center for Geographic Analysis, Harvard University, Cambridge, MA 02138, USA
4 College of Geodesy and Geomatics, Shandong University of Science and Technology, Qingdao 266590, China; liuhongqiang@whu.edu.cn
5 Collaborative Innovation Center of Geospatial Technology, Wuhan 430079, China
6 Key Laboratory of Aerospace Information Security and Trusted Computing, Ministry of Education, Wuhan University, Wuhan 430079, China
* Correspondence: tao.hu@okstate.edu

**Abstract:** A timely understanding of the spatiotemporal pattern and development trend of COVID-19 is critical for timely prevention and control. However, the under-reporting of cases is widespread in fields associated with public health. It is also possible to draw biased inferences and formulate inappropriate prevention and control policies if the phenomenon of under-reporting is not taken into account. Therefore, in this paper, a novel framework was proposed to explore the impact of under-reporting on COVID-19 spatiotemporal distributions, and empirical analysis was carried out using infection data of healthcare workers in Wuhan and Hubei (excluding Wuhan). The results show that (1) the lognormal distribution was the most suitable to describe the evolution of epidemic with time; (2) the estimated peak infection time of the reported cases lagged the peak infection time of the healthcare worker cases, and the estimated infection time interval of the reported cases was smaller than that of the healthcare worker cases. (3) The impact of under-reporting cases on the early stages of the pandemic was greater than that on its later stages, and the impact on the early onset area was greater than that on the late onset area. (4) Although the number of reported cases was lower than the actual number of cases, a high spatial correlation existed between the cumulatively reported cases and healthcare worker cases. The proposed framework of this study is highly extensible, and relevant researchers can use data sources from other counties to carry out similar research.

## 1 Introduction

Since the outbreak of the COVID-19, the prevention and control of the epidemic have rapidly become the focus of social and academic attention (Fu et al., 2020; Guan et al., 2020; J. Liu et al., 2021). As a new infectious disease, measures such as isolation, restriction of crowd activities, and wearing masks have been the most effective emergency prevention and control measures. To achieve targeted and accurate prevention and control, it is necessary to immediately obtain a timely understanding of the spatiotemporal patterns of the epidemic and determine the development trend of the epidemic(Chen et al., 2020; da Silva Corrêa & Perl, 2021).

At present, many scholars have used reported cases data to explore the spatiotemporal characteristics of COVID-19, which includes predicting the inflection point of the disease (Shen et al., 2020; Tang et al., 2020) and detecting its spatial distribution and movement (Jia et al., 2020; Q. Liu et al., 2020; Y. Wang et al., 2021). However, under-reporting of reported cases is a very common phenomenon in fields associated with public health, such as epidemiology and biomedicine (Fernández-Fontelo et al., 2016). In particular, for newly detected infectious diseases such as COVID-19, under-reporting is more likely to occur due to the lack of understanding of the disease and strict diagnostic criteria (Cabaña et al., 2020). It is also possible to draw biased inferences and formulate inappropriate urban prevention and control policies if the phenomenon of under-reporting is not taken into account.

To alleviate the problem of data underreporting, the patient death data are regarded more accurate and are used to conduct research on the phenomenon of under-reporting, such as reconstructing the true epidemic level of the COVID-19 and exploring the impact of under-reporting cases on mortality rate and effective reproduction number (Backer et al., 2020; Lau et al., 2021; Linton et al., 2020; Prado et al., 2020; Saberi et al., 2020). Although the abovementioned studies analyzed the impact of under-reporting on the epidemic situation from multiple perspectives, some discrepancies still remain. First, existing methods mainly estimate the statistical characteristics of COVID-19 from the perspective of time, ignoring the impact of space on the estimation of statistical characteristics. Second, death data still needs to be accurately reported, and there may be a deviation between the number of reported deaths and the actual number of deaths due to COVID-19.

Compared with patient death data, information on healthcare workers with COVID-19 can be considered as a more accurate method for sampling due to a smaller dataset. In the early stage of the epidemic, healthcare worker infections were more easily detected and calculated. Therefore, data on confirmed cases of healthcare workers can more accurately reflect the relevant characteristics of COVID-19 (Gao et al., 2020; Ren et al., 2021; P. Wang et al., 2020). Considering this, a novel framework was proposed to evaluate the spatiotemporal characteristics of COVID-19 outbreak based on the infection data of healthcare workers. The main contributions of this article are as follows:

(1) A novel framework was proposed to explore the impact of under-reporting on COVID-19 spatiotemporal distributions. The proposed framework is highly extensible, and relevant researchers can use data sources from other counties to carry out similar research.
(2) An empirical analysis was carried out based proposed framework using infection data of healthcare workers in Wuhan and Hubei (excluding Wuhan).
(3) An open-source dataset of HCW diagnoses is provided, which both ensures the reproducibility of the study and provides the data needed to support related research on data under-reporting.

## 2 Related Works

In this section, we first reviewed the studies related to the spatiotemporal distribution of COVID-19, and then reviewed the studies related to data under-reporting of COVID-19.

Studies on the spatiotemporal distributions of COVID-19 have mainly focused on reported case data and have explored the patterns and movement of the epidemic, so as to provide scientific basis for relevant measures, such as isolation and the restriction of human activities(Askarizad et al., 2021; Lin et al., 2021). For example, Wang et al. (2021) used scanning statistics to detect the hotspots of new cases each week based on the confirmed cases of COVID-19 at the county level in the United States, thereby characterizing the infection rates during the epidemic. Lak et al. (2021) used spatial regression technology to explore the spatiotemporal spread pattern of 43000 confirmed covid-19 cases at the neighborhood level in Tehran, the capital of Iran. Wang et al. (2021) analyzed the spatiotemporal differences of the spread of the COVID-19 epidemic in 337 prefecture-level cities in China, as well as the social influencing factors and natural influencing factors. Based on mobile phone and confirmed patient data, Jia et al. (2020) developed a spatiotemporal "risk source" model to determine the geographic distribution and growth trends of COVID-19 infections to quickly and accurately assess related risks. Loske (2020) explored the relationship between COVID-19 transmission and transport volumes in food retail logistics by combining transport volume data and confirmed patient data. However, the above studies directly study the spatiotemporal distribution of COVID-19 through reported case data, ignoring the possible under-reporting of data in reported cases (Lau et al., 2021). The conclusions obtained directly from the reported cases may deviate from the actual situation, thereby affecting the judgment of decision-makers(Bastos et al., 2021; Deo & Grover, 2021).

To alleviate the problem of data underreporting, an accurate small sample is used to explore the impact of under-reporting on the estimation of spatiotemporal characteristics (Fellows et al., 2021; Pons-Salort et al., 2021). Additionally, it is common to estimate population characteristics using small samples in the field of statistics (Lauer et al., 2020; Smid et al., 2020). At present, most domestic and foreign researchers regard patient death data as relatively accurate data and use it to study under-reporting (Backer et al., 2020; Fellows et al., 2021; Lau et al., 2021; Linton et al., 2020; Prado et al., 2020). For example, Russell et al. (2020) reconstructed the early global dynamics of under-ascertained COVID-19 cases and infections. Fellows et al. (2021) estimated the incidence, mortality, and lethality rates of COVID-19 among Indigenous Peoples in the Brazilian Amazon. Li et al. (2020) estimated the infection peak time via reported cases data as well as internet search and social media data. Although the above studies considered the underreporting phenomenon, there are still some deficiencies. First, the statistical characteristics of COVID-19 are mostly estimated from the perspective of time, and impact of space on these statistical characteristics is largely ignored. Second, death data still needs to be accurately reported, and there may be a deviation between the number of reported deaths and the actual number of deaths due to COVID-19 (Whittaker et al., 2021).

Therefore, in this paper, infection data of healthcare workers is considered as a more accurate method. Compared with patient death data, healthcare worker infections were more easily detected and calculated (Ren et al., 2021; P. Wang et al., 2020). Additionally, we also proposed a novel framework to explore the impact of under-reporting on COVID-19 spatiotemporal distributions using an accurate small sample.

## 3 Study Area and Data Sources

### 3.1 Study Area

Hubei Province and Wuhan City were the first provinces and cities in China in which COVID-19 was discovered. As of October 2020, the total number of confirmed cases in Hubei Province had reached 68,135, accounting for approximately 81.5% of the total cases in the population, and the total number of confirmed cases in Wuhan had reached 50,340, accounting for approximately 60.80% of the total cases. In order to explore the impact of under-reporting cases on the spatiotemporal distributions of COVID-19, the study area was divided into two parts: Wuhan City and Hubei Province (excluding Wuhan). Hubei Province is located in the central part of China, and Wuhan City is located in the central part of Hubei Province, as shown in Figure 1.

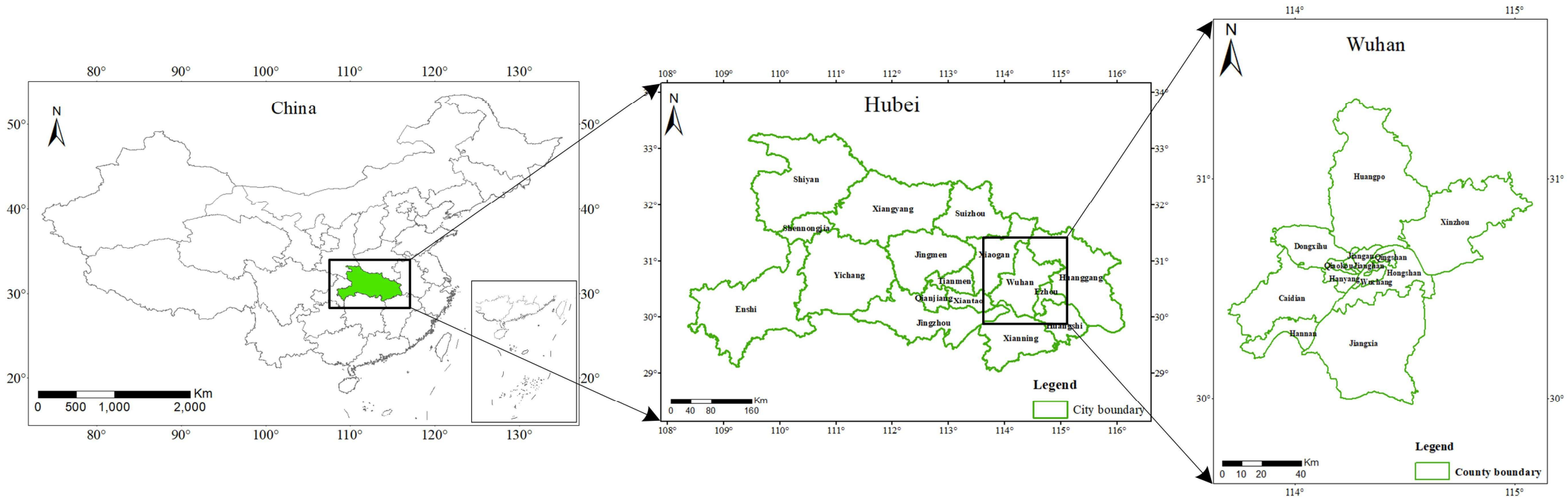

**Figure 1.** Sketch map of the study area.

### 3.2 Data Sources and Data Preprocessing

#### 3.2.1 Data Sources

The data used in this paper included the confirmed cases at the city and county level in Hubei Province and Wuhan, respectively, and information on the healthcare workers in China obtained via retrospective analyses.

The reported cases in Hubei Province were mainly obtained from the Health Commission of Hubei Province (http://www.nhc.gov.cn), and spanned the period from January 15, 2020 to March 31, 2020. The reported cases in Wuhan were mainly obtained from the Wuhan Municipal Health Commission (http://wjw.wuhan.gov.cn), and spanned the period from February 23, 2020 to March 31, 2020.

Data on the confirmed COVID-19 cases of healthcare workers were mainly obtained via retrospective analyses from the Chinese Red Cross Foundation (https://www.crcf.org.cn/), which distributes relief funds to every healthcare worker suffering from COVID-19. As of September 11, 2020, 83 batches of healthcare workers had received foundation assistance. We used crawler technology to obtain 3,743 publications that elucidated the conditions of healthcare workers that suffered from COVID-19. After matching their addresses, information regarding the province, city, and county of all the affected healthcare workers were obtained. The data format is shown in Table 1.

**Table 1.** Sample of confirmed healthcare worker data

| User ID | Date | Province | City | County |
|---|---|---|---|---|
| 1 | 2020-01-15 | Hubei | Wuhan | Huangpo |
| 2 | 2020-01-15 | Hubei | Jingmen | Zhongxiang |
| 3 | 2020-02-04 | Shandong | Qingdao | Shinan |
| …… | …… | …… | …… | …… |
| 3743 | 2020-02-01 | Beijing | Beijing | Xicheng |

3.2.2 Data Preprocessing

Data on the reported cases were collected and processed at the China Data Center and shared on the dataverse platform of Harvard University, which does not require additional data preprocessing(Hu et al., 2020; Yang et al., 2020). Therefore, we mainly conducted data preprocessing for information on healthcare workers suffering from COVID-19, which was obtained via retrospective analyses.

The information on healthcare workers obtained via retrospective analyses is mainly reported to the Chinese Red Cross Foundation in two ways(Chinese Red Cross Foundation, 2020): (1) confirmed information on healthcare workers is directly reported by individuals to the Chinese Red Cross Foundation; and (2) confirmed information on healthcare workers is collected by their respective hospitals, which then report to the Chinese Red Cross Foundation. After examination and approval by the Chinese Red Cross Foundation, the relevant information is published on the website after which the hospital may review it. Those who fail to pass the review do not qualify to be aided. Therefore, data records of unqualified persons should be deleted from the original data. Moreover, the Chinese Red Cross Foundation not only aided infected healthcare workers, but also aided infected or diseased staff during the epidemic; Thus such data records were needed to be deleted. As shown in Figure 2a, after data preprocessing, a total of 3,703 confirmed cases of healthcare workers remained, including 3,655 from Hubei Province and 3,058 from Wuhan City. Compared with the real-time data on the confirmed COVID-19 cases of healthcare workers in the same period (Figure 2b) (Gao et al., 2020), the corresponding data obtained via retrospective analyses exhibit an obvious quantitative advantage.

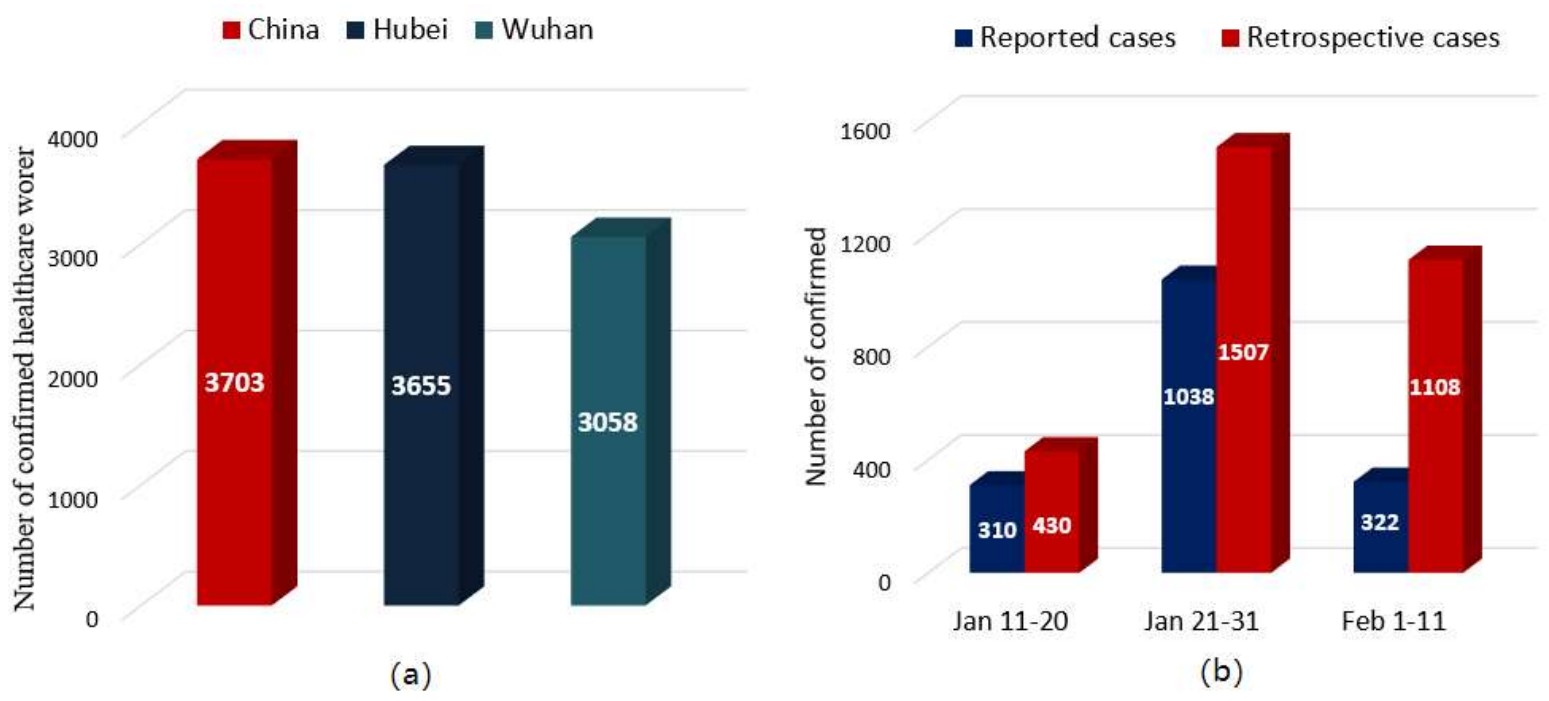

**Figure 2.** Characteristics of healthcare worker infection data. (a) Number of confirmed healthcare workers and (b) number of confirmed cases

## 4 Methods

In this study, the data on confirmed cases of healthcare workers were considered to represent an accurate and small sample space to explore the impact of under-reporting on the spatiotemporal distributions of COVID-19. The overall framework is shown in Figure 3. First, infection inventories were constructed for the healthcare workers and reported cases in Hubei and Wuhan. Second, based on these inventories, the impacts of under-reporting cases on the temporal characteristics were analyzed from three perspectives, namely parameter estimation, temporal correlation, and temporal lag. Finally, the impacts of under-reporting cases on the spatial characteristics were analyzed from the perspective of spatial correlation and spatial lag. This work provides scientific support for researchers that explore the spatiotemporal distribution of COVID-19 using the data on reported cases.

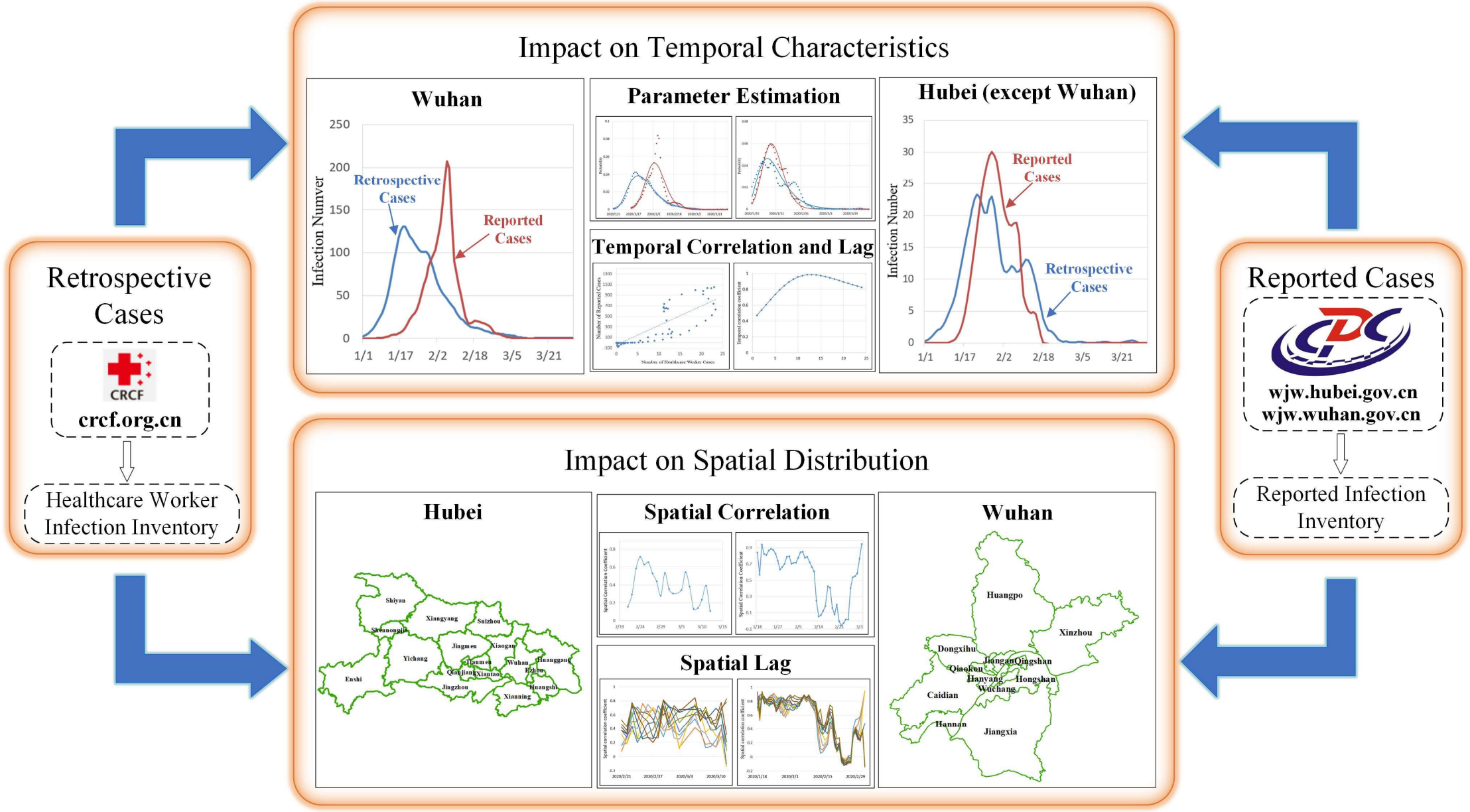

**Figure 3.** Research framework for the impact of under-reporting cases on the spatiotemporal distribution of COVID-19

### *4.1 Construction of Infection Inventory*

Due to the incubation time of the SARS-COV-2 infection, the confirmed time does not represent the infection time of the healthcare workers and reported cases. Therefore, this study constructed a *Healthcare Worker Infection Inventory* and a *Reported Infection Inventory* based on partition statistics from two perspectives: Hubei and Wuhan.

The construction of the *Infection Inventory* can be divided into two steps: (1) Generate a *Confirmed Inventory* based on confirmed healthcare worker data, and (2) Generate an *Infection Inventory* based on the *Confirmed Inventory*. The data structures for the two inventories are shown in Figure 4. *Inventory(S, T)* represents a spatiotemporal dataset, where $S$ and $T$ are the spatial and temporal dimension, respectively; $S = \{s_1, s_2, \cdots, s_m\}$, $T = \{t_1, t_2, \cdots, t_n\}$, $m$ is the total number of spatial objects and $n$ is the total number of timestamps.

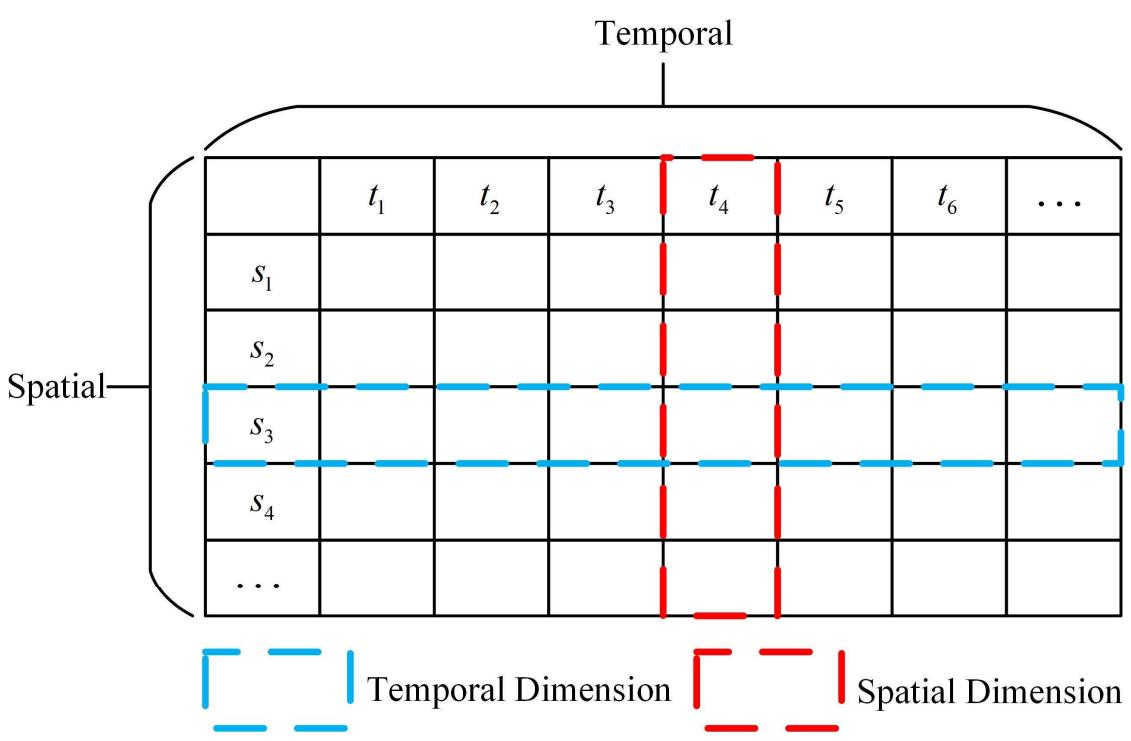

**Figure 4.** Data structure of the confirmed and infection inventories.

Considering the healthcare workers in Wuhan as an example, to construct the *Healthcare Worker Infection Inventory* in Wuhan, the *Confirmed Healthcare Worker Inventory* in Wuhan City was first generated. For example, the calculation method for the number of confirmed healthcare worker on day $t_0$ in Hongshan, Wuhan City, is shown in Equation (1):

$$\begin{cases} H_C_Inventory^{Wuhan}(Hongshan, t_0) = \sum_{i}^{N} |cms_i.County == Hongshan \wedge cms_i.Date == t_0| \\ cms_i \in Retrospective\ Healthcare\ Worker\ Infection\ Set \end{cases}, \quad (1)$$

where $N$ represents the total number of confirmed healthcare worker infections in China, $cms_i$ represents the information of a specific healthcare worker suffering from COVID-19, and $H_C_Inventory^{Wuhan}$ represents the *Confirmed Healthcare Worker Inventory* in Wuhan, which elucidates the changes in confirmed healthcare worker infections in every county of Wuhan over time.

To calculate the *Healthcare Worker Infection Inventory*, we assumed that the incubation time $X$ of SARS-Cov-2 is subject to a lognormal distribution, as seen in other acute respiratory viral infections (Lessler et al., 2009). Lauer et al. (2020) found that the mean and standard deviation of the random variable $ln(X)$ were 1.621 and 0.418, respectively, i.e., $ln(X) \sim N(1.621, 0.418^2)$. Based on the probability distribution function of the incubation time $X$, the daily infected number of healthcare workers can be calculated. For example, the calculation method for the infected number of healthcare workers on day $t_0$ in Hongshan, Wuhan City, is shown in Equation (2):

$$\begin{cases} H_I_Inventory^{Wuhan}(Hongshan, t_0) = \sum_{i}^{n} H_C_Inventory^{wuhan}(Hongshan, t_0 + i) * p_i \\ p_i = \int_{ln(i-1)}^{ln(i)} \frac{1}{\sqrt{2\pi}\sigma} e^{\left(-\frac{(x-u)^2}{2\sigma^2}\right)} \quad i.e., 1 < i < 13 \\ p_i = (1 - \sum_{2}^{12} p_j)/2 \quad i.e., i = 1\ or\ i = 13 \end{cases} \quad (2)$$

where $p_i$ represents the probability that the incubation time is $i$ days; $n$ represents the maximum incubation time (as the incubation time of 98.7% patients is within 13 days, $n$ is set as 13); $u$ represents the mean of lognormal distribution, i.e. 1.621; $\sigma$ represents the standard deviation of lognormal distribution, i.e. 0.418; and $H_I_Inventory^{Wuhan}$ represents the *Healthcare Worker Infection Inventory* in Wuhan. Similarly, the *Reported Infection Inventory* in Wuhan can be expressed as $R_I_Inventory^{wuha}$ , and the corresponding *Healthcare Worker Infection Inventory* and *Reported Infection Inventory* in Hubei can be expressed as $H_I_Inventory^{Hubei}$ and $R_I_Inventory^{Hubei}$, respectively. Taking Hubei Province as an example, Table 2 and Table 3 further show the details of the *Reported Infection Inventory* and *Healthcare Worker Infection Inventory.*

**Table 2.** Sample of reported infection inventory in Hubei

| | **2020-01-24** | **2020-01-25** | **2020-01-26** | **2020-01-27** | **2020-01-28** |
|---|---|---|---|---|---|
| **Wuhan** | 591 | 676 | 839 | 1033 | 1268 |
| **Ezhou** | 40 | 43 | 44 | 43 | 42 |
| **Huanggang** | 141 | 160 | 175 | 191 | 200 |
| **……** | …… | …… | …… | …… | …… |
| **Suizhou** | 59 | 71 | 81 | 90 | 94 |

**Table 3.** Sample of healthcare worker infection inventory in Hubei

| | **2020-01-24** | **2020-01-25** | **2020-01-26** | **2020-01-27** | **2020-01-28** |
|---|---|---|---|---|---|
| **Wuhan** | 106 | 102 | 100 | 101 | 101 |
| **Ezhou** | 5 | 5 | 5 | 4 | 4 |
| **Huanggang** | 5 | 4 | 5 | 5 | 6 |
| **……** | …… | …… | …… | …… | …… |
| **Suizhou** | 1 | 1 | 1 | 1 | 2 |

*4.2 Statistical Model*

The $T$ in $Inventory(S,T)$ is essentially a collection of time-series data which records the temporal characteristics of healthcare worker infection. Studies have shown that counts of the less frequent infections typically follow Poisson distribution, whereas those of more frequent infections may follow approximately normal distribution (Farrington et al., 1996; Unkel et al., 2012). In the early stages of the epidemic, SARS-Cov-2 infections were frequent. Therefore, two samples of healthcare worker cases and reported cases were used to estimate the infection peak time and the infection time interval of the epidemic from three distributions: normal, lognormal, and gamma. The probability density function of normal, lognormal, and gamma are shown in Equations (3), (4), and (5), respectively:

$$Normal(x;u,\sigma) = \frac{1}{\sqrt{2\pi}\sigma} e^{\left(-\frac{(x-u)^2}{2\sigma^2}\right)}, \quad (3)$$

$$Lognormal(x;u_l,\sigma_l) = \frac{1}{\sqrt{2\pi}\sigma_l} e^{\left(-\frac{(lnx-u_l)^2}{2\sigma_l^2}\right)}, \quad (4)$$

$$Gamma(x;\alpha,\beta) = \frac{\beta^\alpha}{\Gamma(\alpha)} x^{\alpha-1} e^{-\beta x}, \quad (5)$$

where $u$ and $\sigma$ represent the fixed parameters of normal, $u_l$ and $\sigma_l$ represent the fixed parameters of lognormal, $\alpha$ and $\beta$ represent the fixed parameters of gamma, and $\Gamma(\alpha)$ represents the gamma function. In this study, the maximum likelihood estimation was used to fit the hyper parameters of the three distributions. As gamma and lognormal are skew distribution, the median of the distribution was used to approximate the peak time of infection.

In addition, the cumulative distribution functions of normal, lognormal, and gamma were used to estimate the time interval of the infection. For example, the calculation method of infection time interval in normal distribution is shown in Equation (6) as follows:

$$\begin{cases} infection\ time\ interval = [\underline{x}, \overline{x}] \\ \underline{x} = F_{normal}^{-1}(0.025) \\ \overline{x} = F_{normal}^{-1}(0.975) \\ F_{normal}(x) = \int_{-\infty}^{x} Normal(t)dt \end{cases} \tag{6}$$

where $F_{normal}(x)$ represents the cumulative distribution functions of normal, $F_{normal}^{-1}(x)$ represents the inverse function of $F_{normal}(x)$, $\underline{x}$ represents lower bound of interval, and $\overline{x}$ represents upper bound of interval. That is, the infection time interval represents the time span of an epidemic infection in 95% of patients. In this study, when the difference between the infection peak and infection interval estimated based on the two samples was small, under-reporting cases had less impact on temporal characteristics, and vice versa.

*4.3 Pearson Correlation Coefficient*

The Pearson's correlation coefficient is used to measure the degree of correlation between two series(Nahler, 2009). As the $Inventory(S,T)$ contains both temporal and spatial dimensions, we calculated the temporal and spatial correlation. As shown in Figure 5, the spatial correlation was obtained by $h_i_t_3$ and $r_i_t_3$, and the temporal correlation was obtained by $h_i_s_3$ and $h_i_s_3$. The temporal and spatial correlation coefficients of the two series were calculated using Equations (7) and (8), respectively:

$$r_{temporal} = \frac{Cov(h_i_s_i, r_i_s_i)}{\sqrt{D(h_i_s_i)}\sqrt{D(r_i_s_i)}} \tag{7}$$

$$r_{spatial} = \frac{Cov(h_i_t_i, r_i_t_i)}{\sqrt{D(h_i_t_i)}\sqrt{D(r_i_t_i)}} \tag{8}$$

where $h_i_s_i$ represents the time series of healthcare worker infections in a specific spatial region, $r_i_s_i$ represents the time series of reported infections in a specific spatial region, $h_i_t_i$ represents the spatial series of healthcare worker infections in a specific time, $r_i_t_i$ represents the spatial series of reported infections in a specific time, $Cov(X,Y)$ represents the covariance of the series $X$ and $Y, D(X)$ represents the variance of the series $X$, and $\sqrt{D(X)}$ represents the standard deviation of the series $X$. The value range of the correlation coefficient is [-1,1]. When the correlation coefficient is greater than 0, the time series $X$ and $Y$ are positively correlated; if it is equal to 1, the time series $X$ and $Y$ are completely positively correlated; if it is less than 0, the time series $X$ and $Y$ would show a negative correlation; and finally, if it is equal to -1, the time series $X$ and $Y$ would show a complete negative correlation. In this study, if the two series showed a significant positive correlation, the under-reporting of cases had less impact on the spatiotemporal characteristics, and vice versa.

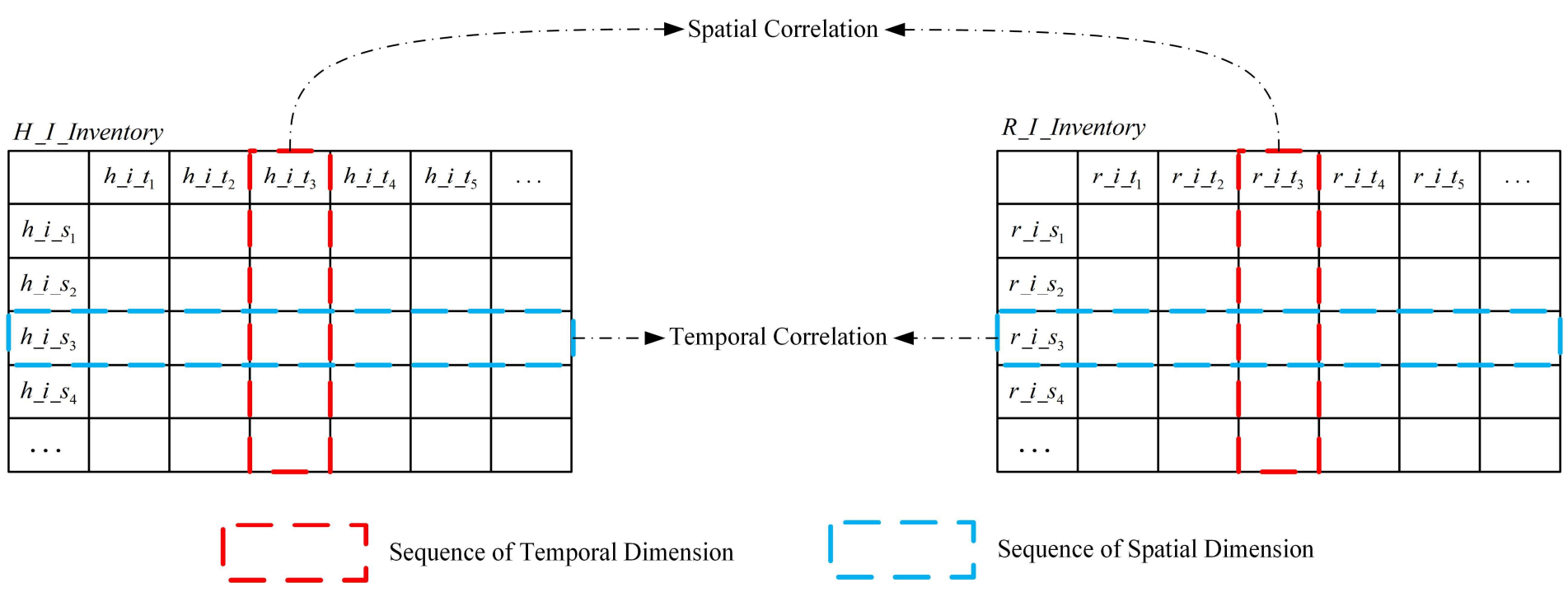

**Figure 5.** Temporal and spatial correlation coefficients

### *4.4 Cross-correlation Function*

The cross-correlation function measures the impact of under-reporting cases on the spatiotemporal characteristics from the perspective of lag. As shown in Figure 6, the cross-correlation function can be understood as a correlation coefficient with a lag.

With regard to the spatial dimension, the spatial lag of series $h_i_t_i$ and $r_i_t_i$ was calculated using Equation (9), as follows:

$$f_{spatial}(\varphi) = \frac{Cov(h_i_t_{i+\varphi}, r_i_t_i)}{\sqrt{D(h_i_t_{i+\varphi})}\sqrt{D(r_i_t_{i+\varphi})}} \quad (9)$$

where $f_{spatial}(\varphi)$ is the spatial correlation coefficient between series $h_i_t_{i+\varphi}$ and $r_i_t_i$ at lag $\varphi$, $r_i_t_i$ represents the spatial series of reported infections in a specific time, $h_i_t_{i+\varphi}$ represents the spatial series of healthcare worker infections lagging $\varphi$ days compared to time $i$.

With regards to the temporal dimension, the temporal lag of series $h_i_s_i = \{h_i_s_i^t\}_{t=1}^n$ and $r_i_s_i = \{r_i_s_i^t\}_{t=1}^n$ was calculated using Equation (10), as follows:

$$\begin{cases} f_{temporal}(\varphi) = \dfrac{\gamma_{temporal}(\varphi)}{\sqrt{\sigma_{h_i_s_i}}\sqrt{\sigma_{r_i_s_i}}} \\ \gamma_{temporal}(\varphi) = E\left[\left(h_i_s_i^t - \overline{h_i_s_i^t}\right)\left(r_i_s_i^{t+\varphi} - \overline{r_i_s_i^{t+\varphi}}\right)\right] \end{cases} \quad (10)$$

where $f_{temporal}(\varphi)$ is the temporal correlation coefficient between series $h_i_s_i$ and $r_i_s_i$ at lag $\varphi$, $h_i_s_i^t$ represents the time series of healthcare worker infections in a specific spatial region, $r_i_s_i^{t+\varphi}$ represents the time series of reported infections lagging $\varphi$ days compared to time $i$ in a specific spatial region.

In the definition, the cross-correlation function can be regarded as a function of the lag, and the lag value that maximizes the cross-correlation function is the average delay time of actual infection. The formal definition can be seen in Equation (11):

$$\begin{cases} \hat{\varphi}_{temporal} = argmax\left(f_{temporal}(\varphi)\right) \\ \hat{\varphi}_{spatial} = argmax\left(f_{spatial}(\varphi)\right) \end{cases} \quad (11)$$

where $\hat{\varphi}_{spatial}$ represents the estimated delay in space, $\hat{\varphi}_{temporal}$ represents the estimated delay in time, the mean of $f_{temporal}(\varphi)$ and $f_{spatial}(\varphi)$ are same as those in Formulas (9) and (10). The smaller the $\hat{\varphi}$, the smaller the impact of under-reporting cases on the spatiotemporal characteristics, and vice versa.

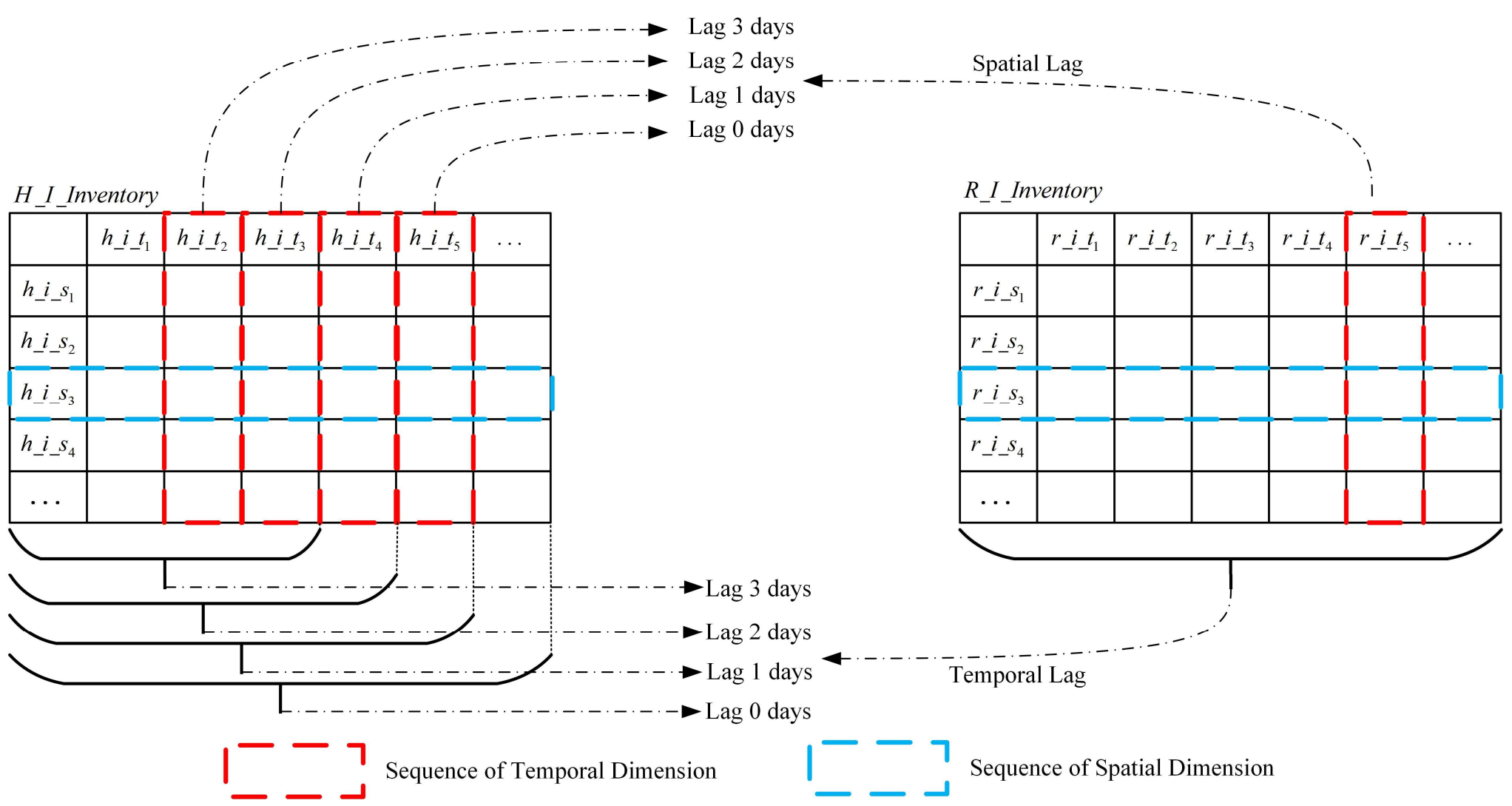

**Figure 6**. Temporal and spatial lag

## 5 Experimental Results and Analysis

### *5.1 Impact of Under-reporting Cases on Temporal Characteristics*

To analyze the impact of under-reporting cases on the temporal characteristics of COVID-19, we first used the data on reported cases and confirmed healthcare worker infections to estimate the epidemic peak and infection interval in Wuhan and Hubei (except Wuhan). Then, we further analyzed the temporal correlation and temporal lag of the data on reported cases and confirmed the healthcare worker infections in Wuhan and Hubei (except Wuhan).

#### 5.1.1 Impact of Under-reporting Cases on Infection Peak and Interval

Figure 7 shows the fitting results of normal, lognormal, and gamma distribution for healthcare worker cases and reported cases in Wuhan, and the mean square error (MSE) is used to evaluate the fitting effect of each distribution. The results showed that the MSE of lognormal distribution was the lowest, which indicated that lognormal distribution was the most suitable to describe the evolution of epidemic with time in Wuhan. In addition, Tables 4 and 5 show the statistical characteristics of healthcare worker cases and reported cases in Wuhan. According to the lognormal distribution, the peak time of infection in Wuhan was estimated to be on February 3 from the data on daily reported cases. However, it was estimated to be on January 24 from the data on daily healthcare worker infection cases, i.e., with a difference of 11 days. The standard deviation estimated from daily reported cases was 7.6 days; 95% of the patients in Wuhan were infected from January 21 to February 19, i.e., within 30.0 days. However, the standard deviation estimated by the daily healthcare worker infection cases was 10.6 days; therefore, 95% of the patients in Wuhan were more likely to be infected from January 8 to February 22, i.e., within 45.6 days.

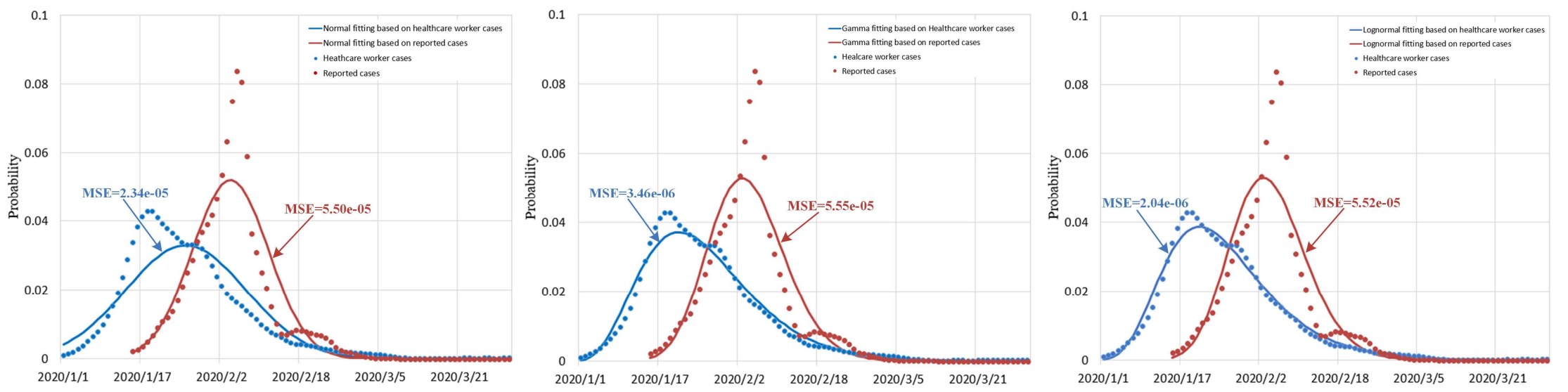

(a) (b) (c)

**Figure 7.** Fitting results of statistical distribution for healthcare worker cases and reported cases in Wuhan: (a) normal distribution, (b) gamma distribution, and (c) lognormal distribution.

**Table 4.** Estimated results of the median and standard deviation for healthcare worker cases and reported cases in Wuhan

| Distribution | | Median (peak) | | Standard Deviation (days) | |
|---|---|---|---|---|---|
| | | Healthcare workers cases | Reported cases | Healthcare workers cases | Reported cases |
| Normal | Estimate | Jan 25 | Feb 3 | 12.1 days | 7.6 days |
| | 95% CI | Jan 24–Jan 26 | Feb 3–Feb 4 | 10.8 days–13.5 days | 7.4 days–8.0 days |
| Lognormal | Estimate | Jan 23 | Feb 3 | 11.8 days | 7.6 days |
| | 95% CI | Jan 23–Jan 24 | Feb 2–Feb 3 | 11.2 days–12.3 days | 7.4 days–7.9 days |
| Gamma | Estimate | Jan 24 | Feb 3 | 11.6 days | 7.6 days |
| | 95% CI | Jan 23–Jan 24 | Feb 3–Feb 4 | 11.1 days–12.1 days | 7.4 days–7.9 days |

**Table 5.** Estimated results of percentiles and infection time intervals of healthcare worker cases and reported cases in Wuhan

| Percentiles | | Healthcare workers cases | | | Reported cases | | |
|---|---|---|---|---|---|---|---|
| | | Normal | Lognormal | Gamma | Normal | Lognormal | Gamma |
| 2.5th | Estimate | Jan 1 | Jan 8 | Jan 7 | Jan 19 | Jan 21 | Jan 20 |
| | 95% CI | Dec 30–Jan 4 | Jan 5–Jan 9 | Jan 6–Jan 8 | Jan 18–Jan 20 | Jan 19–Jan 22 | Jan 19–Jan 22 |
| 97.5th | Estimate | Feb 18 | Feb 22 | Feb 21 | Feb 18 | Feb 19 | Feb 19 |
| | 95% CI | Feb 15–Feb 21 | Feb 20–Feb 25 | Feb 18–Feb 25 | Feb 17–Feb 20 | Feb 18–Feb 21 | Feb 18–Feb 21 |
| Infection time interval | Estimate | 47.5 days | 45.6 days | 45.0 days | 30.1 days | 30.0 days | 30.0 days |
| | 95% CI | 40.6 days–55.0 days | 42.6 days–49.7 days | 40.8 days–49.1 days | 27.6 days–32.3 days | 27.5 days–32.4 days | 27.5 days–32.0 days |

Figure 8 shows the fitting results of normal, lognormal, and gamma distribution for healthcare worker cases and reported cases in Hubei (excluding Wuhan). The results showed that the MSE of lognormal and gamma distribution was relatively small. In addition, Tables 6 and 7 show the statistical characteristics of healthcare worker cases and reported cases in Hubei (excluding Wuhan). Compared with Wuhan, the impact of under-reporting cases on the temporal characteristics of the epidemic in Hubei (excluding Wuhan) was relatively small. According to the lognormal distribution, the peak time of infection in Hubei (except Wuhan) estimated from the data on daily reported cases was January 28, whereas that estimated from the data on daily healthcare worker infection cases was February 29, i.e., only differing by one day. The standard deviation estimated from the daily reported cases was 6.8 days. Further, 95% of the patients in Hubei (excluding Wuhan) were infected from January 17 to February 12, i.e., within 23.7 days. The standard deviation estimated by the daily healthcare worker infection cases was 10.2 days, and therefore, 95% of the patients in Hubei (excluding Wuhan) were more likely to be infected from January 15 to February 23, i.e., within 39.3 days.

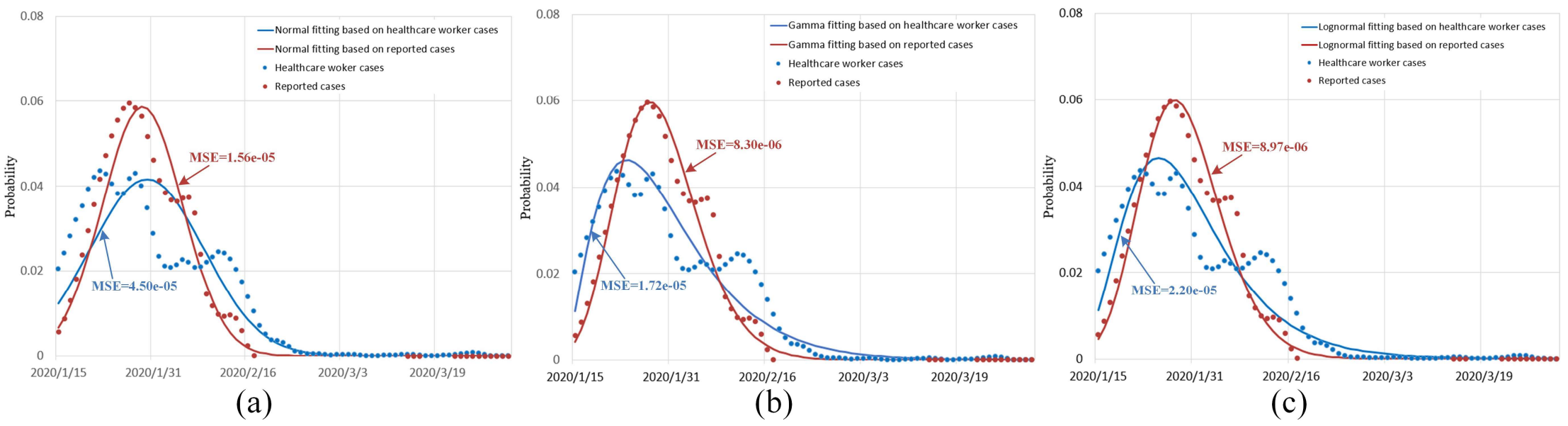

**Figure 8.** Fitting results of statistical distribution for healthcare worker cases and reported cases in Hubei (excluding Wuhan): (a) normal distribution, (b) gamma distribution, and (c) lognormal distribution.

**Table 6.** Estimated results of the median and standard deviation for healthcare worker cases and reported cases in Hubei (excluding Wuhan)

| Distribution | | Median (peak) | | Standard Deviation | |
|---|---|---|---|---|---|
| | | Healthcare workers cases | Reported cases | Healthcare workers cases | Reported cases |
| Normal | Estimate | Jan 29 | Jan 29 | 9.6 days | 6.8 days |
| | 95% CI | Jan 29–Jan 30 | Jan 28–Jan 29 | 9.2 days–10.0 days | 6.4 days–7.1 days |
| Lognormal | Estimate | Jan 28 | Jan 29 | 10.2 days | 6.8 days |
| | 95% CI | Jan 27–Jan 29 | Jan 27–Jan 29 | 9.6 days–10.9 days | 6.4 days–7.2 days |
| Gamma | Estimate | Jan 28 | Jan 28 | 10.4 days | 6.9 days |
| | 95% CI | Jan 26–Jan 28 | Jan 27–Jan 29 | 9.7 days–11.0 days | 6.4 days–7.2 days |

**Table 7.** Estimated results of percentiles and infection time intervals for healthcare worker cases and reported cases in Hubei (excluding Wuhan)

| Percentiles | | Healthcare workers cases | | | Reported cases | | |
|---|---|---|---|---|---|---|---|
| | | Normal | Lognormal | Gamma | Normal | Lognormal | Gamma |
| 2.5th | Estimate | Jan 11 | Jan 15 | Jan 16 | Jan 15 | Jan 17 | Jan 17 |
| | 95% CI | Jan 10–Jan 12 | Jan 14–Jan 16 | Jan 15–Jan 16 | Jan 14–Jan 16 | Jan 16–Jan 18 | Jan 16–Jan 18 |
| 97.5th | Estimate | Feb 17 | Feb 23 | Feb 24 | Feb 11 | Feb 12 | Feb 13 |
| | 95% CI | Feb 16–Feb 19 | Feb 21–Feb 26 | Feb 22–Feb 26 | Feb 10–Feb 12 | Feb 11–Feb 14 | Feb 11–Feb 14 |
| Infection time interval | Estimate | 37.6 days | 39.3 days | 39.3 days | 26.7 days | 26.7 days | 26.7 days |
| | 95% CI | 35.5 days–39.5 days | 36.8 days–41.9 days | 36.7 days–42.1 days | 25.1 days–28.0 days | 25.1 days–28.1 days | 25.1 days–28.1 days |

In general, when the phenomenon of under-reporting is not considered, the infection peak and infection interval estimated by the reported cases may be significantly different from the actual infection peak and infection interval. Among them, the estimated infection peak time may be earlier than the actual infection peak time, and the estimated infection time interval may be smaller than the actual infection time interval. In addition, the impact of under-reporting phenomenon on Wuhan is greater than that on Hubei (excluding Wuhan).

5.1.2 Impact of Under-reporting Cases on Temporal Correlation

Figure 9 shows the temporal correlation between the confirmed healthcare worker cases and the reported cases in Wuhan. The temporal correlation coefficient between the data on daily reported cases and new healthcare worker cases in Wuhan was 0.336, which indicated that the temporal correlation between daily reported cases and actual infection cases was weak in Wuhan. The correlation coefficient between the cumulative reported cases and the cumulative healthcare worker cases in Wuhan was 0.926. As shown in Figure 9d, although there was a strong temporal correlation between cumulative reported cases and actual infection cases in Wuhan, the distance between the trend line and the scatter points was still large in the early stage of the epidemic. This indicated the temporal correlation gradually increased over time, and the phenomenon of under-reporting would gradually decrease over time.

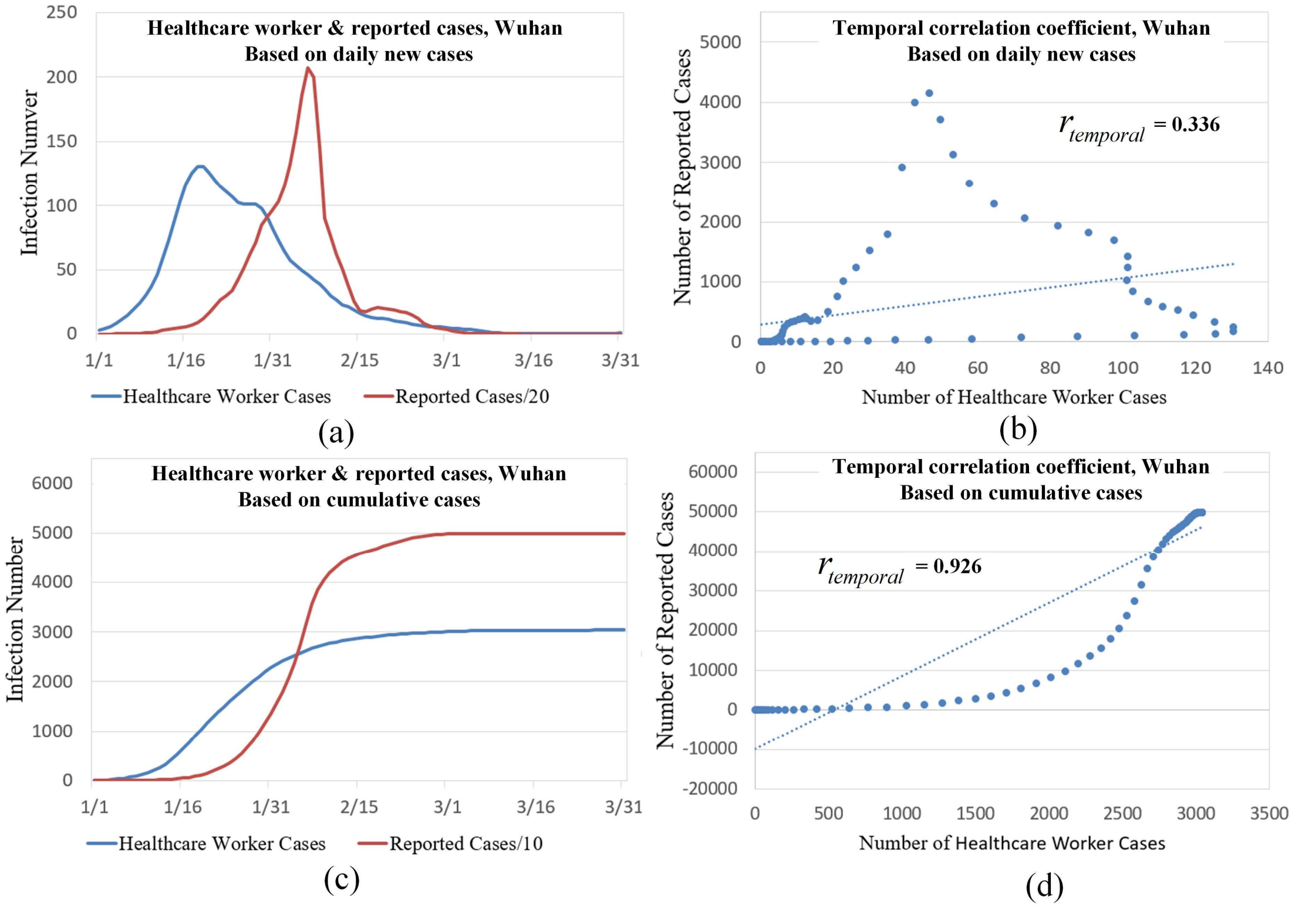

**Figure 9.** Temporal differences between healthcare worker cases and reported cases in Wuhan. (a) Healthcare worker and reported cases in Wuhan, based on daily new cases; (b) temporal correlation coefficient in Wuhan, based on daily new cases; (c) healthcare worker and reported cases in Wuhan, based on cumulative cases; and (d) temporal correlation coefficient in Wuhan, based on cumulative cases.

Figure 10 shows the temporal correlation between the confirmed healthcare worker cases and the reported cases in Hubei (except Wuhan). Compared with the temporal correlation coefficient in Wuhan, the temporal correlation coefficient in Hubei (excluding Wuhan) was significantly improved. For example, the correlation coefficient between the daily reported cases and new healthcare worker cases in Hubei (excluding Wuhan) was 0.828, which indicated that the temporal correlation between daily reported cases and actual infection cases was relatively strong in Hubei (excluding Wuhan). In addition, the correlation coefficient between the cumulative reported cases and the cumulative healthcare worker cases in Wuhan was 0.988, and the scattered points were mostly around the trend line. This indicated that the impact of under-reporting cases on Hubei (excluding Wuhan) was relatively small.

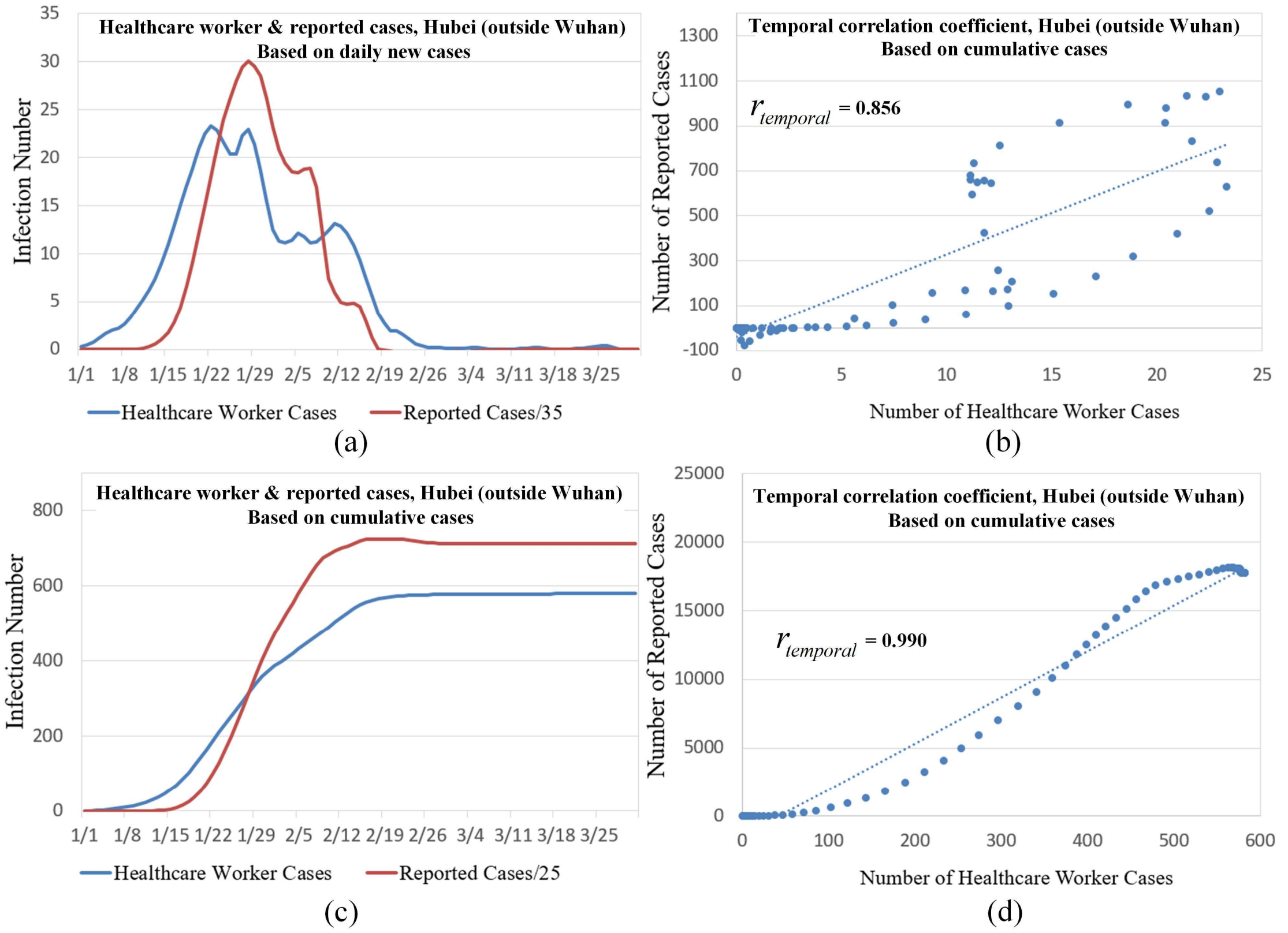

**Figure 10.** Temporal differences between healthcare worker cases and reported cases in Hubei (excluding Wuhan). (a) Healthcare worker and reported cases in Hubei (excluding Wuhan), based on daily new cases; (b) temporal correlation coefficient in Hubei (excluding Wuhan), based on daily new cases; (c) healthcare worker and reported cases in Hubei (excluding Wuhan), based on cumulative cases; and (d) temporal correlation coefficient in Hubei (excluding Wuhan), based on cumulative cases.

### 5.1.3 Impact of Under-reporting Cases on Temporal Lag

According to Figures 9 and 10, if the phenomenon of under-reporting is not considered, the estimated peak time of infection in the reported cases may lead to a time lag. Therefore, we quantitatively analyzed the time lag in Wuhan and Hubei (excluding Wuhan). Figure 11 shows the time-lag results of the healthcare worker infection cases and reported cases in Wuhan and Hubei (excluding Wuhan). With the increase in $\varphi$, the autocorrelation function showed a trend of first increasing and then decreasing, thereby revealing that there was a certain lag in the temporal characteristics in Wuhan and Hubei (excluding Wuhan). When $\hat{\varphi}$ = 13 days, the autocorrelation function of Wuhan provided the maximum value, and the correlation coefficient between the daily reported cases and new healthcare worker cases was 0.984. When $\hat{\varphi}$ = 3 days, the autocorrelation function of Hubei (excluding Wuhan) displayed the maximum value, and the correlation coefficient between the daily reported cases and new healthcare worker cases was 0.972.

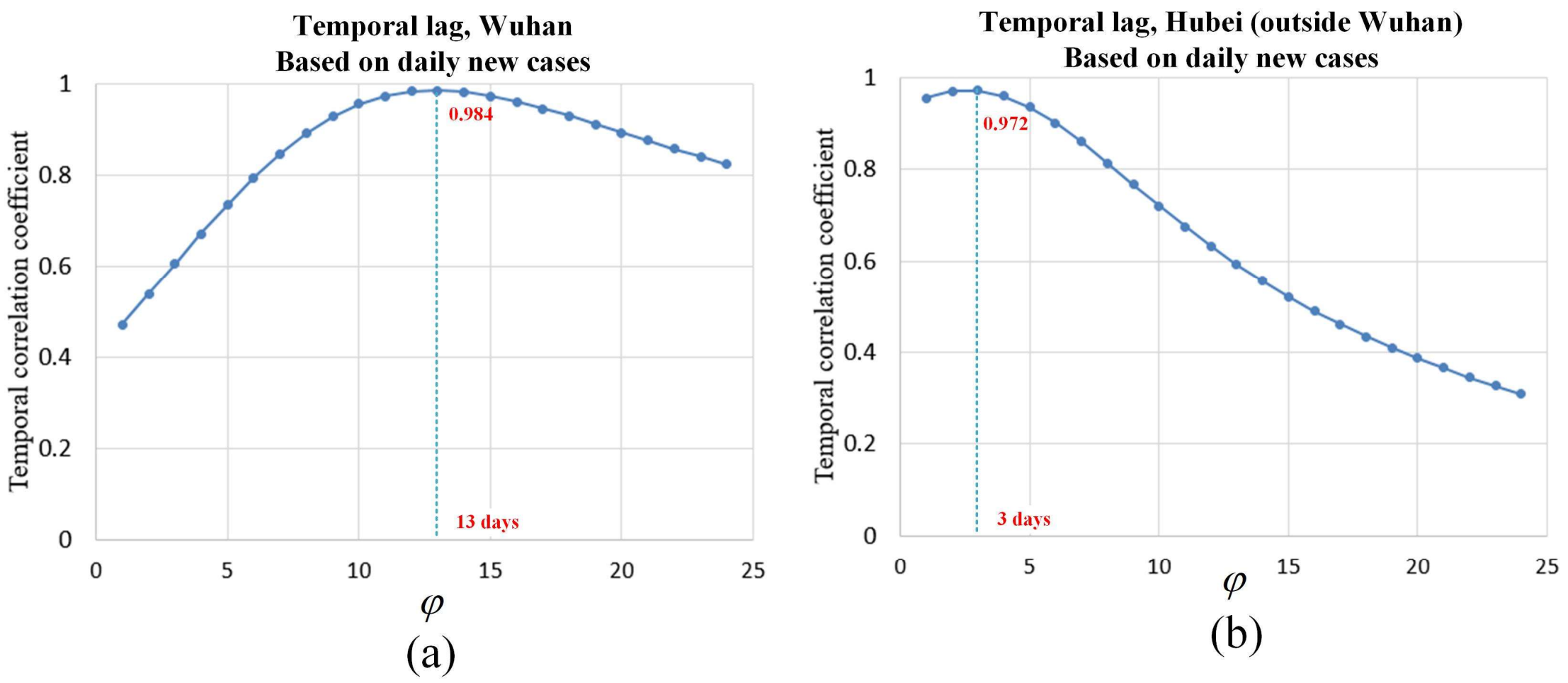

**Figure 11.** Temporal lag between the healthcare worker cases and under-reporting cases: **(a)** Wuhan City and **(b)** Hubei Province (excluding Wuhan).

In addition, we further quantitatively analyzed the time lag phenomenon of the counties in Wuhan and the cities in Hubei (excluding Wuhan). In Hubei, we calculated the time lag in the cities where the cumulative number of healthcare worker infections exceeded 10. In Wuhan, we calculated the time lag in the counties that still reported daily cases after February 21. Figure 12 shows that the time lag phenomenon has spatial heterogeneity in Wuhan and Hubei (excluding Wuhan). Regarding the spatial scale of Hubei, the closer to Wuhan, the greater the time lag there was; for example, the time lag in areas around Wuhan could have been up to 8 days. Regarding the spatial scale of Wuhan, the time lag near the city center was larger as the central area of Wuhan was the most affected region with regard to the number of cases.

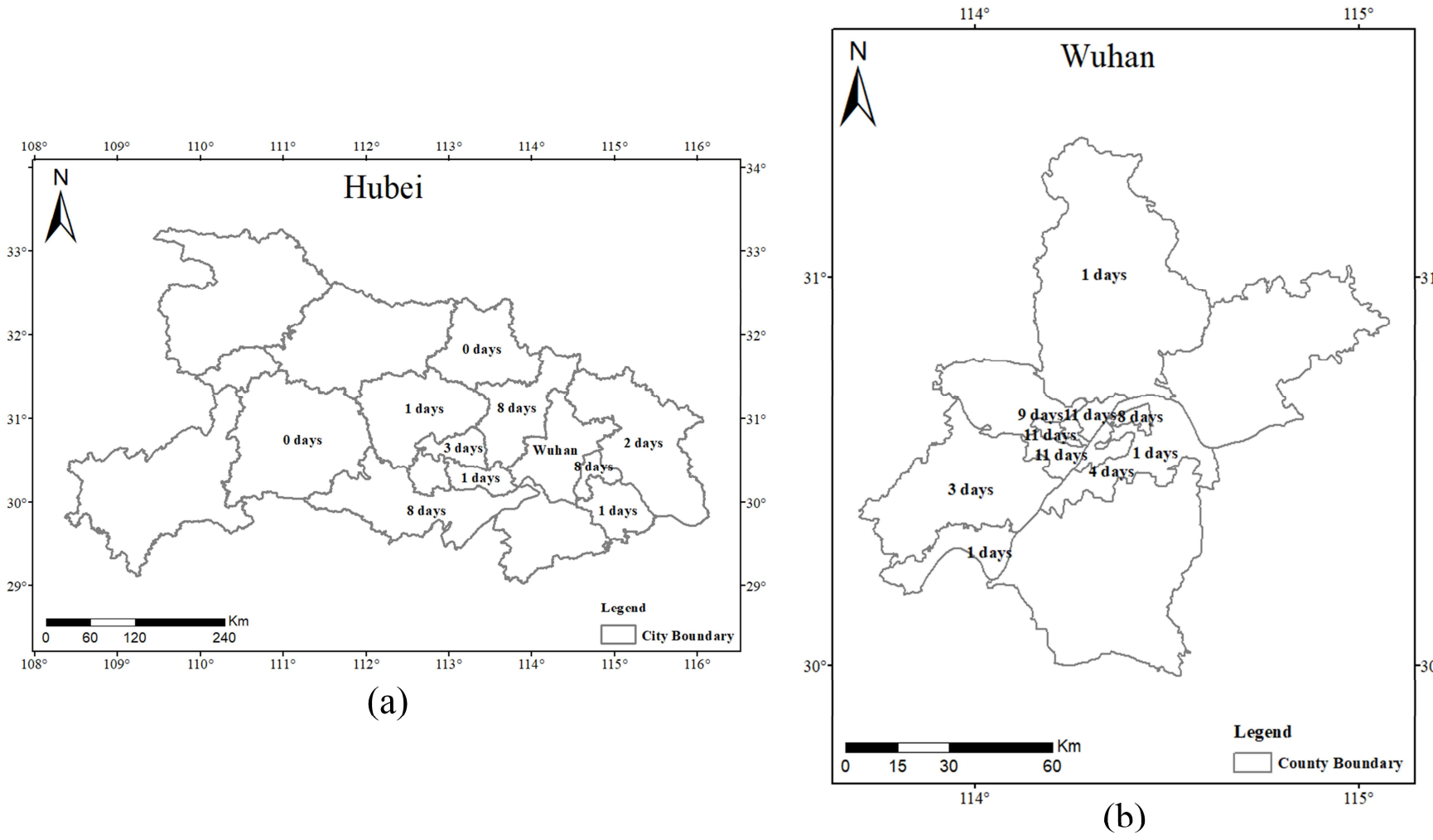

**Figure 12.** Temporal lag of the counties in Wuhan and the cities in Hubei (excluding Wuhan): **(a)** Wuhan City and **(b)** Hubei Province (excluding Wuhan).

In general, the phenomenon of under-reporting has a great impact on the estimated temporal characteristics of the epidemic, and the impact in Wuhan is greater than that in Hubei (excluding Wuhan). According to the time of epidemic occurrence in different regions, the impact of under-reporting in the early onset area was greater than that in the late onset area, and the impact on the early stage was greater than that on the later stage.

*5.2 Impact of Under-reporting Cases on Spatial Characteristics*

In order to analyze the impact of under-reporting cases on the spatial characteristics of COVID-19, we first analyzed the spatial distribution and spatial correlation of the data on reported cases and confirmed healthcare worker infections on a single time node in Wuhan and Hubei (excluding Wuhan). Then, the spatial lag of the data on reported cases and confirmed healthcare worker infections was further analyzed in Wuhan and Hubei (excluding Wuhan).

5.2.1 Impact of Under-reporting Cases on Spatial Correlation

Figure 13 shows the spatial distribution and spatial correlation of the healthcare worker infections and reported cases in Wuhan on March 6. According to Figures 7a and 7b, except for the Jiangxia District, the spatial distribution of healthcare worker infections and reported cases in Wuhan was quite similar, which indicates that, although the level of reported cases was less than the number of actual cases, the spatial distribution of reported cases still reflected the spatial distributions of COVID-19. Moreover, the spatial correlation coefficient between the newly reported cases and the healthcare worker cases was 0.544 in Wuhan on March 6, whereas that between the cumulative reported cases and the healthcare worker cases was 0.889. These results show that there was a certain deviation between the newly reported and healthcare worker cases in Wuhan on March 6; however, there was a high correlation between the cumulative reported cases and healthcare worker cases, which implied that the phenomenon of under-reporting had little impact on the spatial distribution of COVID-19 in Wuhan.

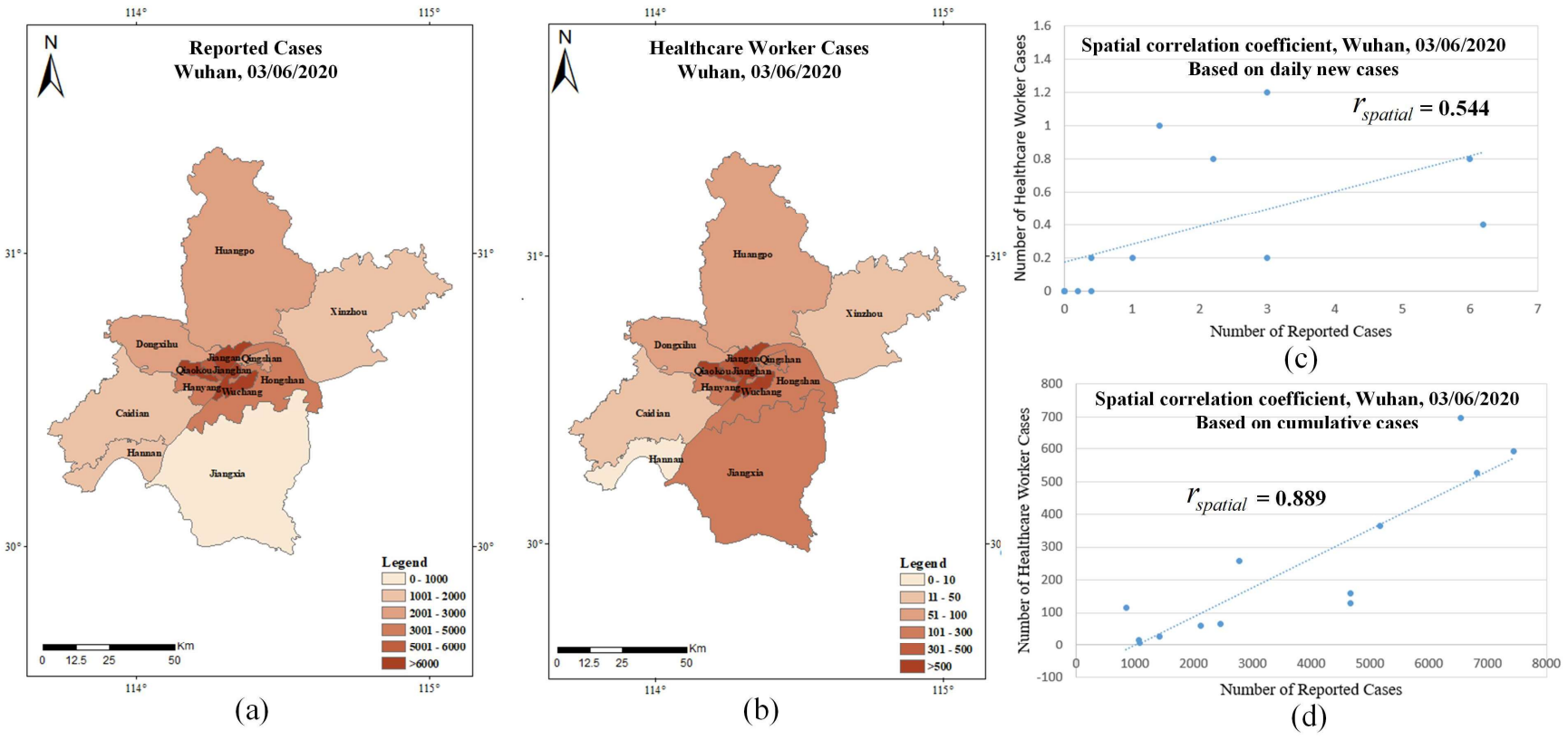

**Figure 13.** Spatial correlations between healthcare worker cases and reported cases in Wuhan. (**a**) Reported cases in Wuhan on March 6, 2020; (**b**) healthcare worker cases in Wuhan on March 6, 2020; (**c**) spatial correlation coefficient in Wuhan on March 6, 2020, based on daily new cases; and (**d**) spatial correlation coefficient in Wuhan on March 6, 2020, based on cumulative cases

Figure 14 shows the spatial distribution and spatial correlation of healthcare worker and reported cases in Hubei (excluding Wuhan) on February 6. Compared with Wuhan, the spatial distributions of healthcare worker and reported cases in Hubei (excluding Wuhan) had a high similarity. Moreover, the spatial correlation coefficient between the newly reported cases and healthcare worker cases was 0.816 in Hubei (excluding Wuhan) on February 6, whereas that between the cumulative reported and healthcare worker cases was 0.846. The results showed that the cumulative reported cases and daily new cases are highly correlated with the healthcare worker infection situation in Hubei (excluding Wuhan), which indicates that the phenomenon of under-reporting had little impact on the spatial distribution of COVID-19 in Hubei (excluding Wuhan).

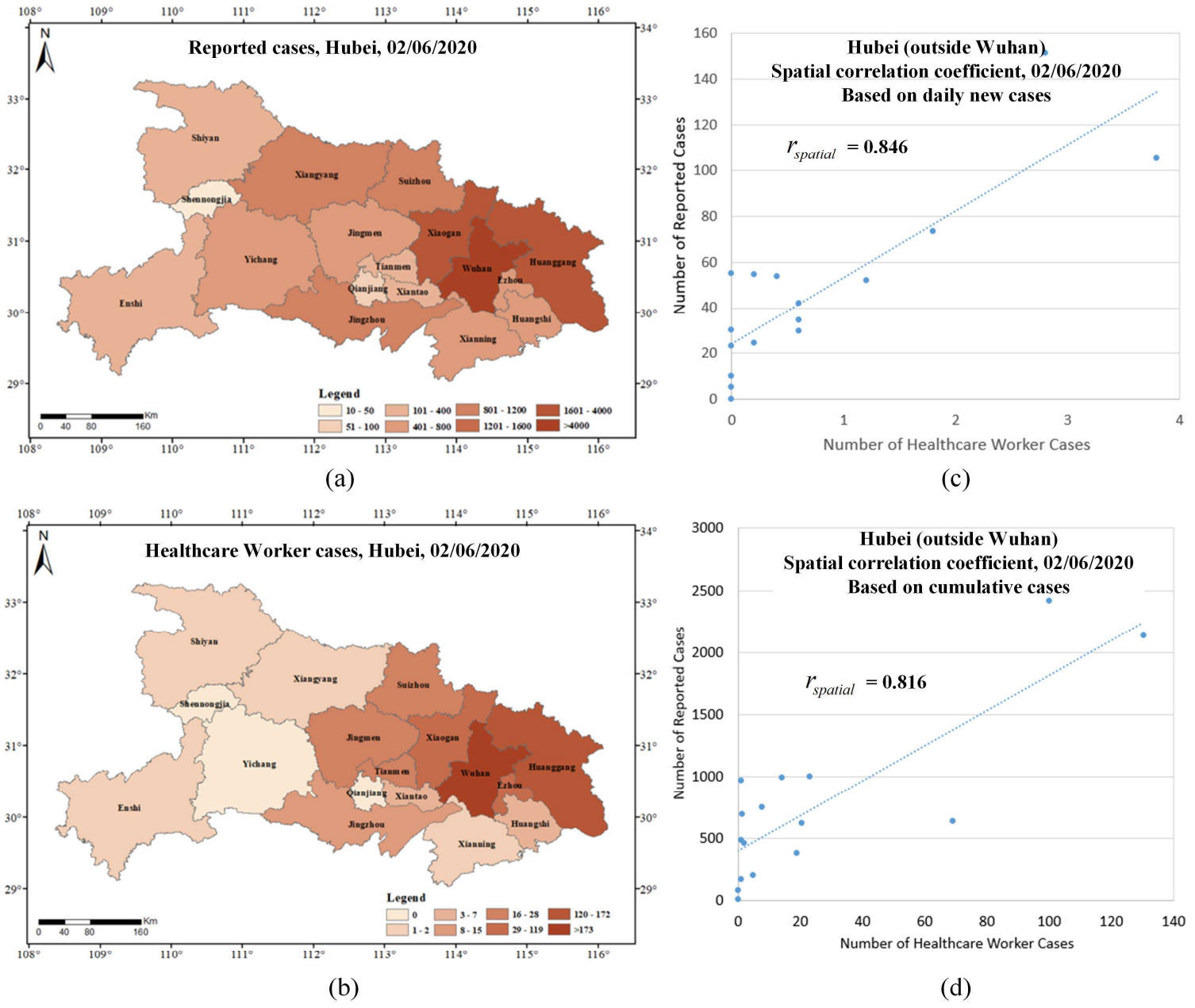

**Figure 14.** Spatial correlation between healthcare worker cases and reported cases in Hubei (excluding Wuhan). (**a**) Reported COVID-19 cases in Hubei (excluding Wuhan) on February 6, 2020; (**b**) healthcare worker cases in Hubei (excluding Wuhan) on February 6, 2020; (**c**) spatial correlation coefficient in Hubei (excluding Wuhan) on February 6, 2020, based on daily new cases; and (**d**) spatial correlation coefficient in Hubei (excluding Wuhan) on February 6, 2020, based on cumulative cases

As Figures 13 and 14 only analyzed the spatial correlation between the healthcare worker and reported cases at a single time node in Wuhan and Hubei (excluding Wuhan), we further demonstrated the spatial correlation changes over time in both locations, and the results are shown in Figure 15.

The results show that the spatial correlation coefficient of daily new cases in Wuhan and Hubei (excluding Wuhan) have a certain volatility, which may be caused by the spatial heterogeneity of temporal lag. Secondly, the spatial correlation of new cases in Hubei (excluding Wuhan) dropped sharply from February 11 to February 29; this is because the relevant department made some corrections to the reported cases data, which led to the decline in the spatial correlation in a short period of time. However, after February 29, the spatial correlation of new cases in Hubei (excluding Wuhan) had rapidly increased and exceeded the correlation coefficient in the early stage of the epidemic.

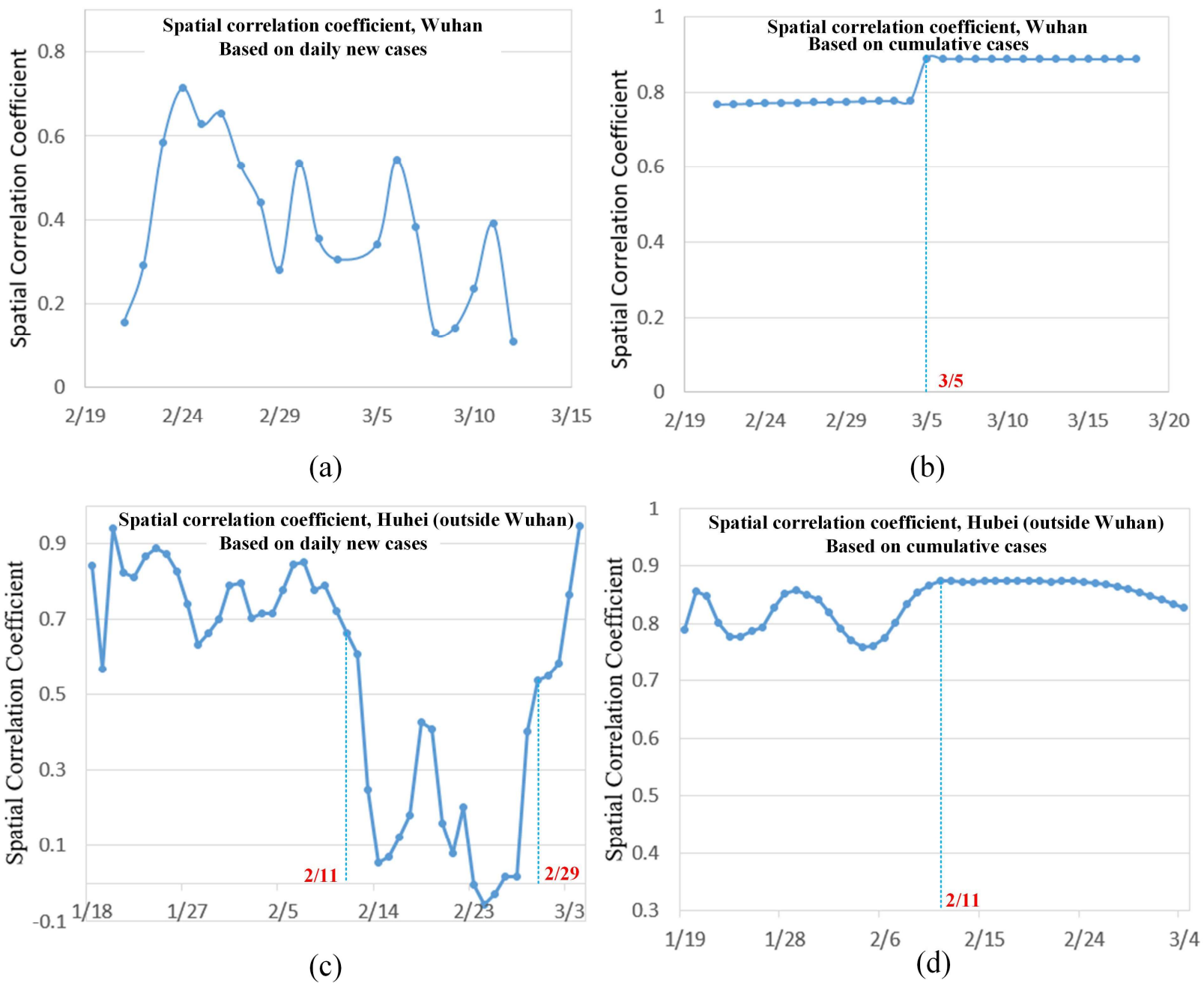

**Figure 15.** Changes in the spatial correlation between healthcare worker cases and under-reporting cases over time. (**a**) Spatial correlation coefficients in Wuhan, based on daily new cases; (**b**) those in Wuhan, based on cumulative cases; (**c**) those in Hubei (excluding Wuhan), based on daily new cases; and (**d**) those in Hubei (excluding Wuhan), based on cumulative cases.

Moreover, the spatial correlation coefficients of the cumulative cases in Wuhan and Hubei (excluding Wuhan) were high. Among them, the spatial correlation coefficient of cumulative cases in Wuhan increased sharply on March 5, as the Wuhan Municipal Government revised the statistical method used to obtain the information on reported cases on March 5 for improving the accuracy of the data. The spatial correlation coefficient of the cumulative cases in Hubei (excluding Wuhan) fluctuated slightly, prior to February 11. This is because in the early stages of the pandemic, the under-reporting phenomenon was more prevalent. As time passes, the phenomenon of under-reporting will be alleviated and the data on reported cases can better reflect the spatial distribution characteristics of COVID-19.

Overall, the impact of the under-reporting phenomenon during the early stages of the pandemic was greater than that during its later stages, which is not only observed with regard to the temporal characteristics, but also in terms of daily new cases. In addition, although the number of reported cases was lower than the number of actual cases, the spatial distribution of the cumulative reported cases in Wuhan and Hubei (excluding Wuhan) still reflected the spatial pattern of COVID-19. Therefore, it is appropriate to use the data on cumulative reported cases to study the spatial distribution characteristics of COVID-19.

#### 5.2.2 Impact of Under-reporting Cases on Spatial Lag

Figure 16 shows the spatial lag of healthcare worker infection and reported cases in Wuhan and Hubei (excluding Wuhan). The results show that the spatial lag of cumulative cases had an insignificant effect on the spatial correlation in Wuhan and Hubei (excluding Wuhan).

However, the spatial lag of daily new cases had a greater effect on spatial correlation, and the effect on Wuhan was greater than that on Hubei (excluding Wuhan). To analyze the spatial lag phenomenon further quantitatively, we fixed the lag days and averaged the corresponding spatial correlation coefficients. Table 8 shows the average spatial correlation coefficient for specific lag days. The results show that the spatial correlation coefficient based on daily new cases in Wuhan increased rapidly with the increase in lag days, whereas that based on daily new cases in Hubei (excluding Wuhan) changed by a small amount with the increase in lag days. This shows that the daily new cases are greatly affected by under-reporting. Among them, Wuhan experienced a great impact due to under-reporting than Hubei (excluding Wuhan). In addition, the spatial correlation coefficient based on the cumulative cases in Wuhan and Hubei (excluding Wuhan) changed by a small amount with the increase in lag days. This indicates that accumulative cases in Hubei (excluding Wuhan) and Wuhan were less affected by the under-reporting phenomenon.

Overall, the impact of under-reporting cases on spatial lag was similar to the impact on spatial correlation. That is, the impact of under-reporting on the early stages of the pandemic was greater than that on its later stages, and its impact on daily new cases was greater than that on accumulative cases.

**Table 8.** The average spatial correlation coefficient in specific lag days

| Lag days | Wuhan | | Hubei (outside Wuhan) | |
|---|---|---|---|---|
| | Daily new cases | Cumulative cases | Daily new cases | Cumulative cases |
| 0 | 0.3897 | 0.8117 | 0.5429 | 0.8414 |
| 1 | 0.3709 | 0.8111 | 0.5429 | 0.8420 |
| 2 | 0.3939 | 0.8106 | 0.5429 | 0.8427 |
| 3 | 0.4165 | 0.8101 | 0.5429 | 0.8439 |
| 4 | 0.4492 | 0.8096 | 0.5362 | 0.8456 |
| 5 | 0.4923 | 0.8091 | 0.5362 | 0.8475 |
| 6 | 0.5249 | 0.8085 | 0.5362 | 0.8489 |
| 7 | 0.5392 | 0.8079 | 0.5636 | 0.8495 |
| 8 | 0.5616 | 0.8072 | 0.5636 | 0.8493 |
| 9 | 0.5766 | 0.8066 | 0.5636 | 0.8486 |

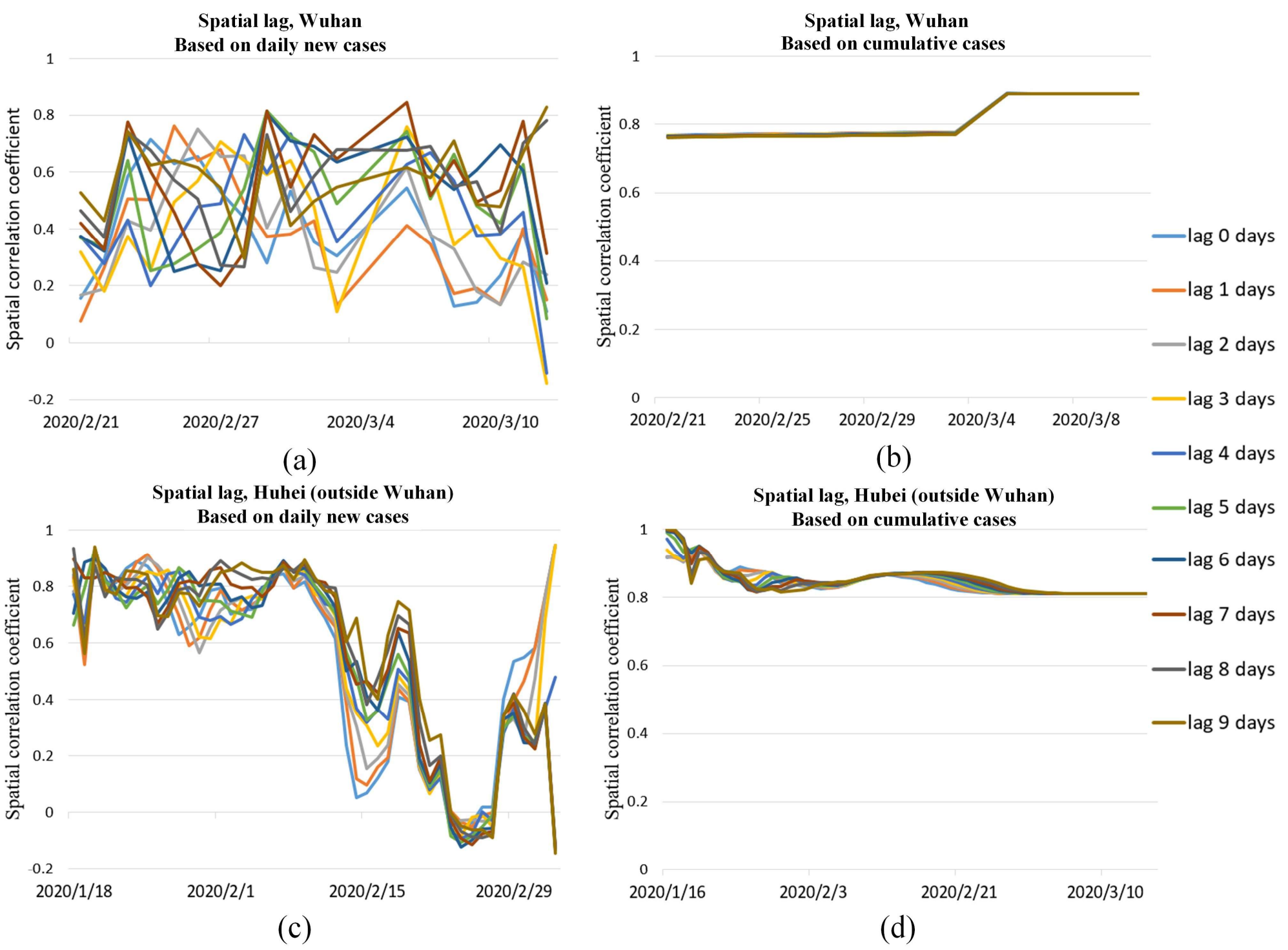

**Figure 16.** (**a**) Spatial lag phenomenon between healthcare worker cases and under-reporting cases in Wuhan, based on daily new cases; (**b**) those in Wuhan, based on cumulative cases; (**c**) those in Hubei (excluding Wuhan), based on daily new cases; and (**d**) those in Hubei (excluding Wuhan), based on cumulative cases.

## 6 Discussions and Conclusions

For COVID-19 epidemic prevention and control, it is important to acquire a timely understanding of the spatiotemporal pattern and determine the development trend of COVID-19 in its early stage. However, under-reporting is a very common phenomenon in public health fields such as epidemiology and biomedicine. When the under-reporting phenomenon is not considered, inaccurate inferences may be produced, which will affect the judgment of decision makers (Paixão et al., 2021). Therefore, in this paper, a novel framework was proposed to explore the impact of under-reporting on COVID-19 spatiotemporal distributions, and empirical analysis was carried out using infection data of healthcare workers in Wuhan and Hubei (excluding Wuhan).

The results show that (1) the lognormal distribution was the most suitable to describe the evolution of epidemic with time; (2) the estimated peak infection time of the reported cases lagged the peak infection time of the healthcare worker cases, and the estimated infection time interval of the reported cases was smaller than that of the healthcare worker cases. (3) The impact of under-reporting cases on the early stages of the pandemic was greater than that on its later stages, and the impact on the early onset area was greater than that on the late onset area. (4) Although the number of reported cases was lower than the actual number of cases, a high spatial correlation existed between the cumulatively reported cases and healthcare worker cases.

According to the results obtained from the proposed framework, the time lag phenomenon should be considered in the time characteristics inferred from the reported cases; otherwise, the urban epidemic prevention and control policies may be unreasonable. Compared with existing methods(Lau et al., 2021; Russell et al., 2020; Shen et al., 2020), the proposed framework does not count the actual number of infections, but treats the healthcare worker infection data as more accurate data to infer the outbreak of the epidemic. In other words, the proposed framework indirectly understands the actual situation of the epidemic, which can avoid complicated calculations and make it more convenient and faster. In addition, the proposed framework of this study is highly extensible. Relevant researchers can not only use data sources from other counties to analyze the impact of under-reported cases on the spatiotemporal distributions of the COVID-19, but also use other types of data sources to analyze the impact of under-reported cases on the spatiotemporal distributions of the COVID-19.

The limitations of this study were as follows: The proposed framework needs more datasets for evaluation. We only used healthcare worker infection data in China to explore the impact of under-reporting cases on the spatiotemporal distributions of COVID-19 and lacks data analysis from other countries. In response to the above limitations, future studies should focus on collecting further domestic and foreign healthcare worker and patient infection data to analyze the impact of under-reporting cases on the spatiotemporal distributions of COVID-19 more accurately and comprehensively.

## Data and codes availability statement

The data and codes that support the findings of this study are available in 'figshare.com' with the identifier: http://doi.org/10.6084/m9.figshare.13560455.

**Funding:** This project was supported by the Key Program of National Natural Science Foundation of China (No. 41830645), Funding program: CAE Advisory Project No. 2020-ZD-16, National Science Foundation, USA (Grant Nos. 1841403, 2027540).

**Acknowledgements:** The authors would like to thank the anonymous referees, editor, and Prof. Shuming Bao (email: sbao@umich.edu) for their helpful comments and suggestions.